\documentclass{JHEP}
\usepackage{amssymb}
\usepackage{bbm,epsfig}

\title{Melvin Matrix Models}

\author{Lubo\v{s} Motl\\
        Department of Physics and Astronomy, Rutgers University\\
        Piscataway, NJ 08855-0849, USA\\
	\quad\\
	and\\
	\quad\\
	Institute for Theoretical Physics, University of California\\
	Santa Barbara, CA 93106-4030, USA\\
        E-mail: \email{motl@physics.rutgers.edu}}

\author{and Author the second}

\abstract{In this short note
we construct the DLCQ description of the flux seven-branes in
type IIA string theory and discuss its basic properties.  The matrix model
involves dipole fields. We explain the relation of this nonlocal matrix
model to various orbifolds. We also give a spacetime interpretation of the
Seiberg-Witten-like map, proposed in a different context first by Bergman
and Ganor, that converts this matrix model to a local, highly nonlinear
theory.}
\preprint{\hepth{0107002}\\HEP-UK-0011}

\begin{document}
\def\eqn#1#2{\begin{equation}#2\label{#1}\end{equation}}
\def\IZ{{\mathbbm Z}}
\def\IR{{\mathbbm R}}
\def\IC{{\mathbbm C}}
\def\IQ{{\mathbbm Q}}

\section{Introduction}

The Melvin solution of Einstein-Maxwell theory \cite{melvinek} has
recently attracted a new wave of attention of string theorists; see
for instance \cite{andyperle}, \cite{costaperle}, \cite{evropsky}. The
axially symmetric universe with magnetic field parallel to the axis (a
flux-string in 4 dimensions) admits a description in terms of Kaluza-Klein
theory.  The corresponding solution of the five-dimensional gravity turns
out to be flat but has nontrivial identifications. Because we want to
study physics of string theory, the corresponding configuration 
has six more dimensions and represents a flux
seven-brane of type IIA string theory and can be rewritten as a flat
metric in eleven-dimensional M-theory 
\eqn{plocha}{ds^2=-dt^2+dy_m dy^m +dz^2+d\rho^2+\rho^2
d\varphi^2+dx_{11}^2}
with nontrivial identifications
\eqn{ztoto}{(t,y_m,z,\rho,\varphi,x_{11})
\equiv (t,y_m,z,\rho,\varphi+2\pi n_2+n_1\beta,x_{11}+2\pi n_1 R),\quad
n_1,n_2\in\IZ.}
Here $t$ and $z$ denote the coordinates parallel to the flux-brane
and 
$y_m=(y_1,y_2,\dots y_6)$ are the extra six stringy dimensions; $x_{11}$
denotes the standard eleventh coordinate compactified on a circle to give
us type IIA string theory (later, when we construct the matrix model, we
will call it $x_9$)
and $(\rho,\varphi)$ are the polar coordinates
of
the two-plane transverse to the flux-brane. Because we deal with a theory
containing fermions, 
the periodicity of $\varphi$ that leads to a completely identical state
should perhaps be written as $4\pi$.
Also $x_{11}$ is a periodic variable but $x_{11}\to x_{11}+2\pi R$ must be
accompanied\footnote{Our variable
$\beta$ is related to $B$ of \cite{costaperle} by
$\beta=2\pi R B$.} by a twist
$\varphi\to\varphi+\beta$. Clearly
$\beta$ is a periodic
variable with period $4\pi$: $\beta+4\pi$ induces an identification
equivalent to the identification computed from $\beta$: we will always
consider $\beta$ from the fundamental interval $(-2\pi,2\pi]$.

Furthermore for $\beta=2\pi$ the operation $\varphi\to\varphi+2\pi$
accounts
for
a change of sign of the fermions and therefore leads to the Scherk-Schwarz
theory with antiperiodic fermions \cite{ss} which was studied in the
context of string theory by Rohm \cite{rohm}. Bergman and Gaberdiel
conjectured that such a compactification of M-theory is dual to type 0A
string theory \cite{begar} in the same way as M-theory on a ``periodic''
circle is a dual of type IIA string theory. A piece of evidence for this
claim was
also shown in \cite{nonsusyorb} where the matrix desciption of the
Scherk-Schwarz compactification of M-theory was constructed: the matrix
string limit \cite{dvvone}-\cite{dvvthree} of this gauge theory was shown
to describe type 0A closed strings in the Green-Schwarz light-cone
variables. Because the matrix model constructed in the present paper
represents a generalization of the matrix model in \cite{nonsusyorb}, we
will discuss the type 0A matrix string limit, too.

We will see in the following subsection that $\beta$ can be interpreted as
the Ramond-Ramond magnetic field on the axis of the flux-brane and that
values $\beta+4\pi k$ are dual to each other for every $k\in\IZ$.
Furthermore Costa and Gutperle \cite{costaperle} argued that such a
configuration is also dual to type 0A theory with magnetic field
$\beta+2\pi+4\pi k'$. We will also present a DLCQ argument for this
conjecture.

\subsection{The language of the Ramond-Ramond flux-branes}

The flat eleven-dimensional solution (\ref{plocha}) with the
identification indicated by (\ref{ztoto}) admits a ten-dimensional
type IIA interpretation. We must first replace $\varphi$ by a new
coordinate
$\tilde\varphi=\varphi-2\pi R\beta x_{11}$ so that the identification
of $x_{11}$
and $x_{11}+2\pi R$ does not affect $\tilde\varphi$. This construction
thus
involves a dimensional reduction along a helix and $\tilde\varphi$ will
play
the role of the polar coordinate in the resulting type IIA picture. (Note
that $\beta$ and $\beta+4\pi k$ lead to different type IIA descriptions.)

The standard identification of the eleven-dimensional Einstein frame
metric $ds_{11}$ and the ten-dimensional type IIA string frame metric
$ds_{10}$, the dilaton field $\phi$ and the Ramond-Ramond one-form
potential ${\cal A}_\mu$ reads
\eqn{slovnik}{ds_{11}^2=e^{-2\phi/3}ds_{10}^2+e^{4\phi/3}
(dx_{11}+{\cal A}_\mu dx^\mu)^2.}
For our particular solution this gives us
\eqn{iia}{ds_{10}^2=\Lambda^{1/2}
\left(-dt^2+ dy_m dy^m +dz^2 +d\rho^2\right)
+\Lambda^{-1/2}\rho^2 d\tilde\varphi^2}
and
\eqn{iiaa}{e^{4\phi/3}=\Lambda=1+(2\pi R \beta\rho)^2,
\quad {\cal A}_{\tilde\varphi}=\frac{2\pi R\beta\rho^2}{\Lambda}.}

\section{Construction of the matrix model}

M(atrix) theory \cite{bfss} can be used to describe physics of
string/M-theory in some
spacetimes with a simple enough asymptotic structure and
with a sufficient number
of large dimensions, using the so-called Discrete Light Cone Quantization
(DLCQ). 

Our model is a $\IZ$ orbifold of the original BFSS matrix model where the
generator of $\IZ$ acts as 
\eqn{jakucinkuje}{X_{9}\to X_{9}+2\pi R,\qquad X_1+iX_2 \to
(X_1+iX_2)e^{i\beta}.}
We will derive the matrix model using the
classical orbifolding procedure. First, we enhance the gauge group $U(N)$
to $U(N\times\infty)$ where $\infty$ indicates points on the circle
$\sigma\in[0,4\pi]$: matrices $X^i,\Pi^i,\theta^i$ also become operators
on the space of complex functions supported on the circle. Then we impose
the restriction that guarantees that the group generated by
(\ref{jakucinkuje}) is identified with a subgroup of the gauge group,
generated by $e^{i\sigma/2}$. For matrices such as $X_{3},\dots X_{8}$
which are left unaffected by (\ref{jakucinkuje}) this simply means that
\eqn{komutuje}{e^{i\sigma/2} \cdot X_i\cdot e^{-i\sigma/2}=X_i,\qquad
i=3,4,\dots 8}
because $X_i$'s transform as adjoint of $U(N\times \infty)$. This equation
implies that $X_i$ commutes with any function of $\sigma$, i.e. is itself
equal to a function of $\sigma$. The matrix elements are proportional to
$\delta(\sigma_m-\sigma_n)$ where $\sigma_m,\sigma_n$ stand for the two
(continuous) indices. Therefore we can forget one of two $\sigma$'s.
Similarly 
\eqn{komutujee}{e^{i\sigma/2} \cdot X_9\cdot e^{-i\sigma/2}=X_9
+2\pi R}
is solved by
\eqn{kovar}{X_9=x_9(\sigma)-4\pi i R\frac{\partial}{\partial\sigma}}
i.e. $X_9$ becomes the covariant derivative of the resulting
$1+1$-dimensional maximally supersymmetric Yang-Mills theory. Apart from
the general $N\times N$ matrix-valued function of $\sigma$, $X_9$ contains
the term with matrix elements $-4\pi i R\delta'(\sigma_m-\sigma_n)$. And
what about $Z=X_1+iX_2$? The condition reads
\eqn{rotace}{e^{i\sigma/2} \cdot Z\cdot e^{-i\sigma/2}=Z e^{i\beta}}
which is solved by $Z$ proportional to the factor
$\delta(\sigma_m-\sigma_n+2\beta)$. In the language of the
$1+1$-dimensional Yang-Mills theory, the scalar field $Z$ is not local
anymore: it transforms as $({\bf N}, {\bf \bar N})$ under two $U(N)$
groups located at different points $\sigma_m,\sigma_n$, separated by the
fixed interval $2\beta$. More generally the fields in the Yang-Mills
theory
become dipoles whose length is equal to $2\beta J_{12}$ where
$J_{12}$ is a component of the angular momentum. Namely the spinors
$\theta$ become dipoles of length $\pm\beta$.
% where the sign is proportional to the eigenvalue of $J_{12}$.

\subsection{Type 0A matrix string limit}

For $\beta=2\pi$ all the fields $X_i$ become local fields again (the
length of the dipoles is a multiple of $4\pi$) but all $\theta$'s behave
as dipoles of length $\pm 2\pi$, stretched between the opposite points of
the circle. One can make this theory local by identifying points $\sigma$
and $\sigma+2\pi$. The gauge group is then $U(N)\times U(N)$ -- each
factor comes from one of the opposite points and the two factors get
interchanged if one goes around the reduced circle $\sigma\in[0,2\pi]$.
The fields $X_i$
still transform in 
the adjoint representation of the gauge group while the fermions $\theta$
transform in the bifundamental representation $({\bf N}, {\bf \bar N})$. 

One can derive the matrix string limit \cite{dvvone}-\cite{dvvthree} of
this model. For long strings the dipole character of the fields becomes
irrelevant and the main difference from the type IIA matrix string theory
is the existence of the gauge transformations satisfying (in the nonlocal,
dipole language)  $U(\sigma+2\pi) =-U(\sigma)$, for example
$U(\sigma)=\exp(i\sigma/2)$ under which the bifundamental representation
is odd while the bosonic fields in adjoint do not transform at all. The
Green-Schwarz fermions $\theta$ are
therefore also allowed to be antiperiodic on the long string. Together
with the corresponding GSO-like projection 
(indentified with the requirement of the gauge invariance under $U$)
such a construction gives the
correct description of type 0A string theory using the Green-Schwarz
fermions. Although such a derivation is purely classical and at quantum
level the spacetime interpretation breaks down \cite{nonsusyorb}, one can
still consider
this type 0A matrix string limit to be formal evidence for the
conjecture of Bergman and Gaberdiel \cite{begar} relating type 0A string
theory and the Scherk-Schwarz compactification of M-theory.

In a similar fashion, one can also take the limit of M-theory on an
infinitely small two-torus with periodic conditions on one circle and the
Scherk-Schwarz conditions on the other circle. Perturbatively, such a
model can be understood as an orbifold of type IIB matrix string theory
\cite{dvvone}, \cite{dvvtwo}, \cite{setkind} and leads to type 0B matrix
string theory written in Green-Schwarz variables \cite{nonsusyorb}. The
duality symmetry $SL(2,\IZ)$ of type IIB string theory is reduced down to
its subgroup $\Gamma(2)$ that preserves the boundary conditions.

Quantum mechanically, such nonsupersymmetric configurations are unstable.
The spacetime is not static and we are not justified to quantize the
theory in DLCQ because the spacetime does not contain two null Killing
vectors. At the level of the matrix model, this problem manifests itself
as a two-loop divergence that destroys the spacetime interpretation of the
matrix model \cite{nonsusyorb}. 

However, we can construct supersymmetric
versions of the fluxbranes, too.  The simplest case is to add an extra
twist to (\ref{jakucinkuje}), namely
\eqn{jakdva}{X_3+iX_4 \to
(X_3+iX_4)e^{-i\beta}.}
The complex scalar field $Z'=X_3+iX_4$ will then transform as a dipole
oriented in the opposite direction than $Z=X_1+i X_2$. Such a
``supersymmetric F5-brane'' has been discussed in \cite{andyperle}. It can
be understood as a pair of intersecting F7-branes: each twist represents a
single F7-brane and we try to describe their superposition. Our matrix
model for such an intersection has two complex scalar dipole fields of
length $2\beta$. One
half of the fermions transform as dipoles of length $2\beta$, too. The
other half transform as local fields in adjoint of $U(N)$ and represent
the unbroken supersymmetries. We believe that such a matrix model for the
supersymmetric F5-brane should have the usual spacetime interpretation.

%For a rational twist $\beta=4\pi \cdot p/q$ we obtain a matrix model that
%differs from the matrix model for the $A_{q-1}$ singularity by the
%permutation of the $U(N)$ factors only.

\subsection{The case of the rational twists}

One can of course formulate the matrix model for the flux seven-brane
in the local ``quiver''
language \cite{kvivr}
whenever $\beta$ is a rational multiple of $4\pi$,
$\beta=4\pi\cdot p/q$. The circumference of the new $\sigma$-circle is
then $4\pi/q$ and the dipole fields transform in the representations
derived from a loop quiver diagram (the extended Dynkin diagram of
$A_{q-1}$, i.e. in the bifundamental
representations $({\bf N}_i, {\bf \bar N}_{i+p})$ where $i=1,\dots q$
of the gauge group $U(N)^q$. For $p=1$ and $q\to\infty$ the dipole
character of the fields could be neglected and the theory would be very
similar in spirit to the theory of the artificial dimensions recently
constructed by Arkani-Hamed, Cohen and Georgi \cite{georgi} but we will
not study the details of this correspondence here.

We can also note some properties of the matrix model for the
supersymmetric F5-brane (a pair of intersecting F7-branes).
For a rational twist $\beta=4\pi \cdot p/q$ we obtain a matrix model that
differs from the matrix model with the same gauge group $U(N)^q$
for the $A_{q-1}$ singularity by the 
permutation of the $U(N)$ factors only. This permutation of the factors
associated with $\sigma\to\sigma+4\pi/q$ is the matrix description of the
Ramond-Ramond Wilson line that distinguishes the F5-brane from the
$A_{q-1}$ singularity \cite{andyperle}.

\subsection{Seiberg-Witten map and its interpretation}

The transformation rule of the dipole fields (of length $2\beta$, such
as $Z$, stretched between $\sigma-\beta$ and $\sigma+\beta$) under
the gauge transformations is identical to that of the open (untraced)
Wilson lines
\eqn{ouwl}{{\cal W}_{\sigma-\beta,\sigma+\beta}=P\,
\exp\left(i\int_{\sigma-\beta}^{\sigma+\beta}
A_\sigma(\sigma')d\sigma'\right).}
It is therefore easy to redefine the dipoles in the following way:
\eqn{seiwitmap}{Z(\sigma-\beta,\sigma+\beta)=
P\,
\exp\left(i\int_{\sigma-\beta}^{\sigma}
A_\sigma(\sigma')d\sigma'\right)
\tilde Z(\sigma)
P\,
\exp\left(i\int_{\sigma}^{\sigma+\beta}
A_\sigma(\sigma')d\sigma'\right)}
We chose a symmetric convention where the new $\tilde Z$ sits in the
middle of the Wilson line. We could have defined the map in a less
symmetric way but all such choices lead to a theory with local fields and
infinitely many higher derivative terms. An important thing to note is
that while $\beta$ and $\beta+4\pi k$, $k\in\IZ$ lead to the same physics
(as seen in 11 dimensions), the redefinitions (\ref{ouwl}) 
associated with them are different:
this is the matrix realization of the ``different'' type IIA descriptions
where $\beta$ differs by multiples of $4\pi$.
Our notation $\tilde Z$ coincides with the angular variable
$\tilde\varphi=\varphi-2\pi R\beta x_{11}$ defined earlier: $\tilde Z=
\tilde \rho e^{i\tilde\varphi}$.

This redefinition of the variables is a dipole counterpart of the
Seiberg-Witten map \cite{switten} which was written using the 
open Wilson lines by Liu \cite{swittenliu}.
\eqn{liusw}{F_{\mu\nu}(k)=\int d^D k\,\,
L_*\left[
\sqrt{\det(1-\theta\hat F)}
\left(
\frac{1}{1-\hat F\theta}
\hat F\right)_{\mu\nu}
W(x,C)
\right]e^{ik\cdot x}}
In this formula, the commutative $U(1)$ field strength $F_{\mu\nu}$
is expressed in terms of a straight Wilson line $W(x,C)$. The determinant
and the rational function of the noncommutative $\hat F$ are understood as
a power series expansion. The length of the Wilson line is proportional to
the momentum, $\Delta x^\mu=\theta^{\mu\nu}k_\nu$. The symbol $L_*$
guarantees that all the operators are inserted in a path-ordered fashion.

In fact, Liu's prescription for the Seiberg-Witten map is a natural
generalization of Ganor and Bergman's form of the corresponding map
removing the dipoles. In both cases, the fields are redefined by an open
Wilson line. In the case of dipoles, the Feynman vertices acquire a phase
linear in momentum, and consequently the open Wilson line has a fixed
length. In the noncommutative case, the Feynman vertices include phases
bilinear in momenta, and therefore the length of the Wilson line is
proportional to the momentum.

\section{Conclusions and open questions}

In this short note, we constructed the matrix model for the flux
seven-branes and their intersections. The matrix model involves dipole
fields. Using open Wilson lines, such dipole fields can be converted to
the local fields. The Wilson lines wrapped around the circle $n$ times
lead to different type IIA interpretations of the configuration where the
Ramond-Ramond field strength on the axis differs by $n$ times its period.
The Wilson lines that make all the bosonic fields local while the
fermionic fields transform as dipoles stretched between the opposite
points represent different type 0A interpretations of the theory.

Matrix models with dipoles could be useful to learn something about
physics of strings in backgrounds with a Ramond-Ramond field strength.
Dipole fields can also serve as a simple toy model for noncommutative
geometry. In both cases the Feynman vertices acquire an extra phase; in
the case of the dipoles, the phase is linear in momenta, while in the case
of noncommutative geometry it is bilinear. In both cases, a Seiberg-Witten
map involving open Wilson lines can be used to transform the action into a
very nonlinear dynamics of local fields.

\acknowledgments

I would like to thank the ITP, Santa Barbara, and the ITP workshop on
M-theory for support and hospitality during the progress of this work. 
This work was supported in part by the Department of Energy, grant DOE
DE-FG02-96ER40559. I would also like to thank Andy Strominger, Tom Banks,
Michal Fabinger, Ruth Britto-Pacumio, David Berenstein,
and Joe Polchinski for useful discussions.

\end{document}